# Disorder-induced ordering in gallium oxide polymorphs

Alexander Azarov[1,*], Calliope Bazioti[1], Vishnukanthan Venkatachalapathy[1,2],

Ponniah Vajeeston[3], Edouard Monakhov[1], and Andrej Kuznetsov[1,*]

[1] *University of Oslo, Department of Physics, Centre for Materials Science and Nanotechnology, PO Box 1048 Blindern, N-0316 Oslo, Norway*

[2] *Department of Materials Science, National Research Nuclear University, "MEPhI", 31 Kashirskoe Hwy, 115409 Moscow, Russian Federation*

[3] *University of Oslo, Department of Chemistry, Centre for Materials Science and Nanotechnology, PO Box 1033 Blindern, N-0315 Oslo, Norway*

Polymorphs are common in nature and can be stabilized by applying external pressure in materials. The pressure/strain can also be induced by the gradually accumulated radiation disorder. However, in semiconductors, the radiation disorder accumulation typically results in the amorphization instead of engaging polymorphism. By studying these phenomena in gallium oxide we found that the amorphization may be prominently suppressed by the monoclinic to orthorhombic phase transition. Utilizing this discovery, a highly oriented single-phase orthorhombic film on the top of the monoclinic gallium oxide substrate was fabricated. Exploring this system, a novel mode of a lateral polymorphic regrowth, not previously observed in solids, was detected. In combination, these data envisage a new direction of research on polymorphs in $Ga_2O_3$ and, potentially, for similar polymorphic families in other materials.

Basic principles of polymorphism in crystals are clear: the lattices adapt to the minimum energy in respect with the temperature and pressure. Meanwhile, controllable stabilization of the metastable phases, e.g. at room temperature, is of great interest because it opts for new properties; however, preferably without applying colossal external pressures [1]. For that matter, reaching metastable conditions via ion beam assisted processes with displaced atoms provoking strain accumulation is an interesting alternative. However, the ion-induced disorder may also result in the amorphization, as it typically occurs in such semiconductors as Si [2] or SiC [3]. Even for so-called "radiation hard" semiconductors, e.g. GaN [4] or ZnO [5], the ion irradiation still ends up with high disorder level or partial amorphization.

Importantly, $Ga_2O_3$ is well known for its polymorphism [6] and several attempts were recently performed to study the influence of the ion irradiation on the phase stability in $Ga_2O_3$. For example, Ander *et al*. have detected κ-phase in the β-$Ga_2O_3$ matrix upon heavy ion implants [7], even though the evolution of this process as a function of the implantation parameters has not been explored. More recently, interesting strain accumulation was observed in ion implanted α-$Ga_2O_3$ [8] and β-$Ga_2O_3$ [9]. However, the systematic understanding of the disorder-induced ordering in $Ga_2O_3$ polymorphs was missing in literature. In the present paper, we use a combination of theoretical and experimental approaches to study the response of β-$Ga_2O_3$ to ion irradiation exploring a wide range of the experimental parameters, such as type of implanted ions, accumulated dose and irradiation temperature. As a result, we demonstrate a controllable ion-beam-induced phase engineering resulting in the formation of the highly oriented κ-phase film on the top of the β-$Ga_2O_3$ wafer. This breakthrough paves the way for a new synthesis technology and further explorations of the metastable polymorph films and regularly shaped interfaces in the device components not achievable by conventional thin film deposition methods. Thus, the present study, in addition to its fundamental value, may accelerate the technological development of $Ga_2O_3$, which is one of the most intensively studied ultra-wide bandgap semiconductor right now [10].

In the present study, (010) and (-201) oriented β-$Ga_2O_3$ single crystal wafers (Tamura Corp.) were implanted with different ions, specifically $^{58}Ni^+$, $^{69}Ga^+$, and $^{197}Au^+$. Note that most of the data are shown for systematic $^{58}Ni^+$ ion implants, while $^{69}Ga^+$ and $^{197}Au^+$ results are used for comparison to demonstrate the general character of the phenomena. For that reason, the ballistic defect production rates (without accounting for non-linear cascade density effects [11]) for $^{69}Ga^+$ and $^{197}Au^+$ implants were normalized to that of $^{58}Ni^+$ ion implanted with 400 keV in a wide dose range of $6 \times 10^{13}$ - $1 \times 10^{16}$ cm$^{-2}$. The implantations were performed at room temperature (if not indicated otherwise) and 7° off the normal direction.

The samples were analyzed by a combination of Rutherford backscattering spectrometry in channeling mode (RBS/C), x-ray diffraction (XRD) and scanning transmission electron microscopy (STEM) combined with electron energy loss spectroscopy (EELS). The RBS/C was performed using 1.6 MeV $He^+$ ions incident along [010] direction and 165° backscattering geometry. The XRD 2theta measurements were performed using Bruker AXS D8 Discover diffractometer using Cu $K_{\alpha1}$ radiation in locked-coupled mode.

The STEM and EELS investigations were conducted on an FEI Titan G2 60-300 kV at 300 kV with a probe convergence angle of 24 mrad. The simultaneous STEM imaging was conducted with 3 detectors: high-angle annular dark field (HAADF) (collection angles 101.7-200 mrad), annular dark field (ADF) (collection angles 22.4-101.7 mrad) and annular bright field (ABF) (collection angles 8.5-22.4 mrad). The resulting spatial resolution achieved was approximately 0.08 nm. EELS was performed using a Gatan Quantum 965 imaging filter. The energy dispersion was 0.1 eV/channel and the energy resolution measured using the full width at half maximum (FWHM) of the zero-loss peak was 1.1 eV. Electron transparent TEM samples with a cross-sectional wedge geometry were prepared by mechanical polishing with the final thinning performed by Ar ion milling and plasma cleaning.

The total lattice energies of the $Ga_2O_3$ polymorphs were calculated applying density functional theory (DFT), as implemented in the VASP code [12]. The interaction between the core (Ga:[Ar] and O:[He]) and the valence electrons were described using the projector-augmented wave (PAW) method [13]. The Perdew, Burke, and Ernzerhof (PBE) [14] gradient corrected functional was used for the exchange–correlation part of the potential for the structural optimization. The large energy cut-off of 600 eV was used to guarantee basis-set completeness. The atoms were deemed to be relaxed when all atomic forces were less than 0.02 eV Å$^{-1}$ and the geometries were assumed to get optimized when the total energy converged to ≤1 meV between two consecutive geometric optimization step. The crystal lattice parameters for all $Ga_2O_3$ phases in the equilibrium were computed accordingly and the pressure was applied both isotropically and uniaxially.

Fig.1 summarizes the data proving the ion-beam-induced β-to-κ phase transformation in $Ga_2O_3$ combining (a) XRD, (b) RBS/C, and (c) STEM measurements of (010) β-$Ga_2O_3$ samples irradiated with $^{58}Ni^+$ ions. As it is clearly seen from Fig. 1(a) the virgin β-$Ga_2O_3$ wafer is characterized by a strong reflection around 60.9° attributed to the (020) planes of β-$Ga_2O_3$. The low dose implants (6×10$^{13}$ and 2×10$^{14}$ Ni/cm$^2$) result in the formation of the shoulder peak with some oscillations at the high-angle side of the (020) peak. These shoulder peak/oscillations can be attributed to the accumulation of the compressive strain [15] and has been discussed in ion implanted (010) β-Ga$_2$O$_3$ [9]. However, further dose increase ($1\times10^{15}$ Ni/cm$^2$) releases this strain and leads to the formation of the broad diffraction peak around 63.4°, which shifts to 63.7° for the dose of $1\times10^{16}$ Ni/cm$^2$. Importantly, for this highest dose, the strain shoulder peaks practically disappear and the (020) diffraction peak resembles that in the virgin sample. The diffraction peaks centered at 63.4° and 63.7° were interpreted as signatures of the κ-Ga$_2$O$_3$ (330) and (060) planes, respectively [16], as systematically analyzed in Fig.S1 in Supplementary Materials-I [17]; notably, Fig.S2 illustrates the evolution in the similarly implanted (-201) β-Ga$_2$O$_3$ samples for comparison.

Further insight into the mechanisms of the formation and the integrity of the new phase in Ga$_2$O$_3$ can be performed by channeling analysis. Fig.1(b) shows that for the lowest dose sample, the RBS/C spectrum is characterized by the well resolved Gaussian-like damage peak centered close to the maximum of the nuclear energy loss profile ($R_{pd}$=125 nm according to the SRIM simulations [24]). The magnitude of this peak is well below the amorphous level that is equivalent to the height of the random spectrum in Fig.1(b). The increase of the ion dose up to $2\times10^{14}$ cm$^{-2}$ results in the formation of the box-shape disorder layer reaching ~90% of the random signal. Further dose increase, to $1\times10^{15}$ Ni/cm$^2$, broadens the disordered layer while the disorder level saturates corroborating with the data for β-Ga$_2$O$_3$ obtained previously [25]. Spectacularly, the backscattering yield decreases for the dose of $1\times10^{16}$ Ni/cm$^2$, implying the lattice ordering along the [010] direction, in agreement with the XRD data revealing (060) κ-phase diffraction peak. Thus, it can be concluded that the $Ga_2O_3$ amorphization is suppressed by the β-to-κ phase transition occurring in the implanted region.

To verify the status of this film we performed STEM investigations and Fig. 1(c-h) summarizes the STEM data for the sample implanted with $1\times10^{16}$ Ni/cm$^2$. Specifically, Fig.1(c) shows an ABF-STEM image and strain contrast reveals the formation of two distinct regions – the film and the substrate – of the initially homogeneous β-$Ga_2O_3$ wafer. Selected area electron diffraction (SAED) patterns taken from the unimplanted and implanted regions, i.e. Figs.1(d) and 1(e), illustrate a prominent transformation from monoclinic β- to ordered orthorhombic κ-phase. This phase transformation extends to ~300 nm from the surface and stops abruptly, forming a sharp interface with the β-phase wafer/substrate, see Fig.1(c). The contrast associated with defects/strain inside the κ-$Ga_2O_3$ film gradually increases towards the κ/β interface, see Fig.1(c). However, fast Fourier transforms (FFTs) from high-resolution images taken at the interfacial area (g) and the upper part (h) of the implanted region show that the ordered orthorhombic phase is retained through the depth of the film as compared with the FFT at the β-$Ga_2O_3$ substrate, see Fig.1(f). Thus, SAED and FFTs show the formation of a single-phase ordered orthorhombic κ-phase region both at meso and nano-scale. The sharp spots, in addition to the lack of extra reflections and/or striking of the main reflections, indicate a highly-oriented crystalline film with no signs of high-angle mis-orientations, high-density of mis-oriented grains or amorphization (see also Fig.S3 in Supplementary Materials II [17]). However, minor mis-orientations between adjacent grains cannot be excluded, playing a role in the high de-channeling yield in RBS/C and the broadening of the XRD peaks in Fig.1.

Note that, the implanted region in the low dose sample contains only point defects and defect clusters with no indication of the κ-phase (see Supplementary Materials III [17]). Moreover, the comparison between EELS spectra in Fig.2 provides additional arguments. Indeed, because of different atomic coordination in β- and κ-phases we detected characteristic changes in the fine structure of the EELS spectra acquired in STEM-mode, by comparing low and high-dose implanted samples. In particular, the oxygen K-edge is characterized by two main peaks, labeled A and B in Fig.2, related to the O 2p - Ga 4s and O 2p - Ga 4p bonding, respectively. As seen from Fig.2, prominent changes in the A/B intensity ratio ($I_A/I_B$) occur when the phase transition takes place. Specifically, $I_A/I_B$ decreases in the κ-phase. This can be attributed either to the increase in O 2p – Ga 4p hybridization or to the transfer of electrons from O 2p – Ga 4p band into another band [7,26].

It should be emphasized that the phenomenon of the β-to-κ phase transition is generically related to the accumulation of the lattice disorder and not to the chemical nature of the implanted ions. This was proved by performing control implants with Ga and Au ions (see Figs.S1 and S5 in the Supplementary Materials I and IV [17]).

Further, we investigated the thermal stability of the formed κ-$Ga_2O_3$ film. The RBS/C data show that the anneals at 300-700 °C result in the reduction of the κ-film thickness as revealed by the shifts of the RBS/C spectra in the range corresponding to the β/κ interface region (Fig.3(a)). However, increasing the temperature to 700 °C enhances the yield in the near surface region, indicating the disintegration of the κ-phase film. The corresponding XRD data in Fig.3(b) show that the (060) κ-phase peak position shifts too and for 300-500 °C it moves towards better fits with literature data [16]. This shift may indicate an improvement of the κ-Ga$_2$O$_3$ phase crystallinity because of annealing of the misoriented grains and relaxation of the enhanced strain (see Supplementary Materials V [17]) at the β/κ interface region consistently with the RBS/C data in Fig.3(a).

Moreover, based on the data in Fig.3, lateral solid-phase regrowth of the polymorphs can be emphasized. The kinetics of this process resembles that of the solid phase epitaxial regrowth of the amorphous phase induced by ion irradiation, e.g. for amorphous silicon [2], even though in reality Fig.3 reveals the crystalline phase transition. The inset in Fig. 3(a) plots the rate of this process as a function of the annealing temperature, revealing the activation energy of 0.16±0.1 eV. Notably, in accordance with literature [27] κ-Ga$_2$O$_3$ deposited with conventional techniques starts to degrade at ~650 °C, but the nucleation of the β-phase occurs in the bulk of the κ-phase layer. Thus, our results suggest an important role of the β/κ interface for the thermal stability of radiation-induced κ-Ga$_2$O$_3$ films.

The fundamental reason of the κ-phase metastability is of a thermodynamic origin and we analyzed Ga$_2$O$_3$ phases using the DFT. Fig.4 shows the total energy for α-, β-, and κ-Ga$_2$O$_3$ lattices subjected to isotropic or uniaxial pressure, plotted as a function of volume per formula unit (f.u.) for two different intervals in panels (a) and (b). Notably, there are more Ga$_2$O$_3$ polymorphs available, however α-, β-, and κ-phases are the lowest in energy as seen from Fig.S7 in Supplementary Materials-VI [17]. In particular, the κ-phase exhibits its energy minimum closest to that of β-phase, making the β-to-κ transition likely to occur as soon as sufficient strain is provided (~2%, corresponding to the volume of ~51.35 Å$^3$/f.u. in Fig.4 for isotropic pressure/strain). Thus, we believe that such strain conditions are gradually reached by the high dose implants, finally resulting in the κ-phase film. Meanwhile, the strain sufficient to ignite β-to-α phase transition is only slightly higher, see Fig.4(a); meaning that the formation of the α-phase is also possible, as was also experimentally demonstrated under high isotropic pressure [28]. To start this process, there is an additional activation volume to overcome (as compared to β-to-κ transition), since the energy minima for the α-phase corresponds to the volume of ~50.5 Å$^3$/f.u. in Fig.4(a). This explains the reason of the single-phase κ-film in Fig.1(c) instead of the α/κ-phase mixture or α-phase alone. Thus, the scenario behind the observations in Figs.1 and 2 is consistent the theoretical predictions of the radiation disorder-induced strain reaching a threshold to ignite the β-to-κ transition and, consequently, relaxing the strain by moving the system to the κ-phase energy minimum in Fig.4(a).

Notably, Fig.4 also compares the isotropic and uniaxial pressures, since at present we do not have sufficient experimental arguments to conclude on the strain distribution. Non-isotropic strain conditions are realistic because of the preferential localizations of the radiation defects and building defect complexes with specific configurations [29], likely compressing/stretching certain crystal directions. Nevertheless, for the pressures applied, non-negligible energy difference, as compared with the isotropic strain, occurs in κ-$Ga_2O_3$ only, see Fig.4(b). Interestingly, it shifts the κ-$Ga_2O_3$ energy curve towards the "triple point" intersection for α, β, and κ-phases at the volume of ~51.2 Å$^3$/f.u. in Fig.4(b). Additional important parameter not taken into account in Fig.4 is temperature. Indeed, the data in Fig.4 represent the ground state only, while the temperature shifts the stability range for all phases as illustrated in Fig.S11 in Supplementary Materials-VI [17]. Moreover, temperature might affect the balance between migration and annihilation of the radiation-induced defects [9], making it to an important factor for the localization of the β/κ interface in respect with the $R_{pd}$ region (see Fig.S12 in Supplementary Materials –VII [17]).

In conclusion, we have studied ion-induced phase transitions in $\beta$-Ga$_2$O$_3$. The mechanism involves the lattice disorder-induced strain accumulation that ignites the transition upon reaching sufficient strain threshold as predicted by theory. As the result, we fabricated a [010] oriented orthorhombic $\kappa$-phase film on the top of the [010] oriented monoclinic $\beta$-phase substrate, demonstrating sharp $\kappa/\beta$ interface not previously realized by conventional thin film deposition methods. We also observed a novel mode of solid-phase polymorphic regrowth of $\beta$-phase on the expense of $\kappa$-phase film maintained across the lateral $\kappa/\beta$ interface and exhibiting the activation energy of 0.16±0.1 eV at 300-500°C. In combination, the present work paves the way for further development of the polymorph films and interfaces by ion beam fabrication, envisaging this new research direction for Ga$_2$O$_3$ and, potentially, for similar polymorphic families in other materials.

**ACKNOWLEDGMENTS**

This work was performed within the Research Centre for Sustainable Solar Cell Technology (FME SuSolTech, project number 257639) co-sponsored by the Research Council of Norway and industry partners. The Research Council of Norway is acknowledged for the support to the Norwegian Micro- and Nano-Fabrication Facility, NorFab, project number 295864, and to the Norwegian Center for Transmission Electron Microscopy, NORTEM, project number 197405/F50. We also acknowledge the Research Council of Norway for providing the computer time, project numbers NN2875k and NS2875k) at the Norwegian Supercomputer Facility. The INTPART Program at the Research Council of Norway, projects No. 261574 and No. 322382, enabled the international collaboration.

Phys. Rev. Lett. **128**, 015704 (2022)
Doi: https://doi.org/10.1103/PhysRevLett.128.01570427. I. Cora, Zs. Fogarassy, R. Fornari, M. Bosi, A. Rečnik, and B. Pécz, *In situ* TEM study of κ→β and κ→γ phase transformations in $Ga_2O_3$, Acta Materialia **183**, 216 (2020).

28. D. Machon, P. F. McMillan, B. Xu, and J. Dong, High-pressure study of the β-to-α transition in $Ga_2O_3$, Phys. Rev. B. **73,** 094125 (2006).

29. M. E. Ingebrigtsen, A. Yu. Kuznetsov, B. G. Svensson, G. Alfieri, A. Mihaila, U. Badstübner, A. Perron, L. Vines, and J. B. Varley, Impact of proton irradiation on conductivity and deep level defects in β-$Ga_2O_3$, APL Mater. **7**, 022510 (2019).14

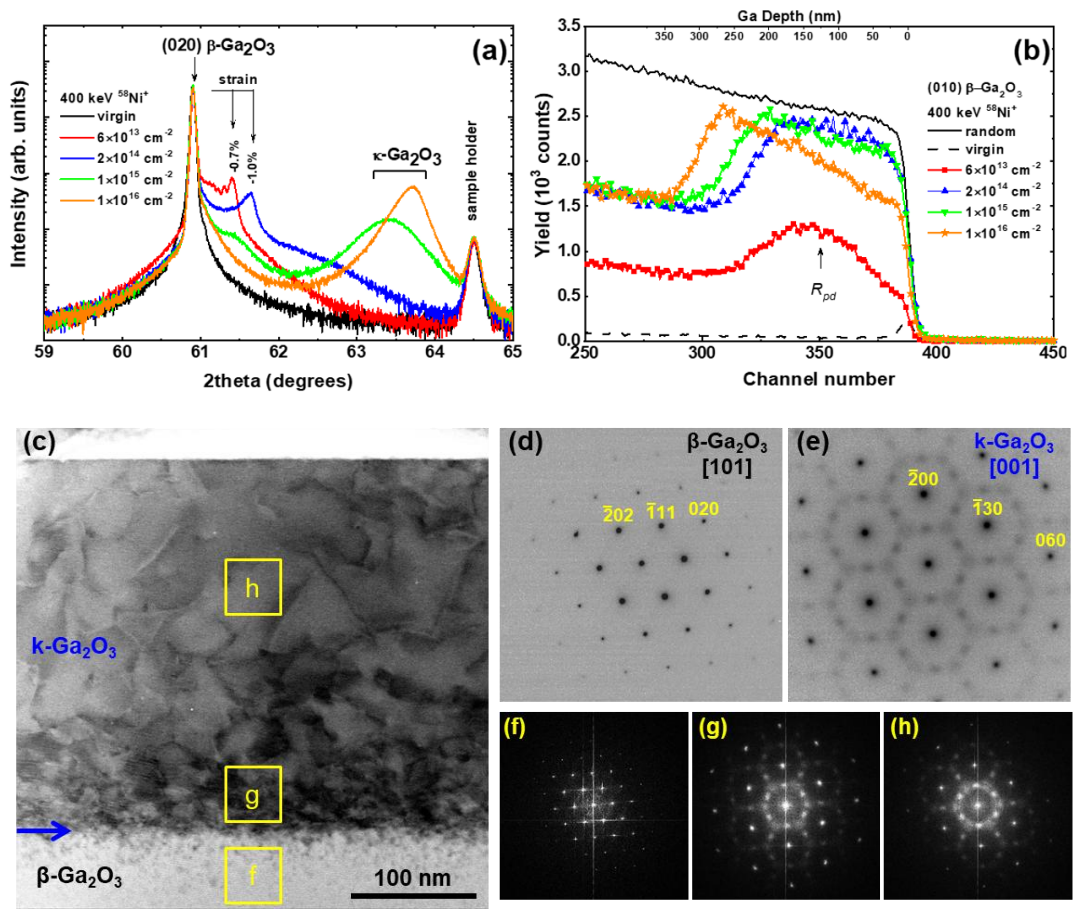

**Figure 1.** Ion-beam-induced β-to-κ phase transformation in $Ga_2O_3$. (a) XRD 2theta scans of the samples implanted with 400 keV $^{58}Ni^+$ ions as a function of ion dose (b) The corresponding RBS/C spectra with the random and virgin (unimplanted) spectra shown for comparison. The maximum of nuclear energy loss profile ($R_{pd}$) predicted with SRIM code [24] simulations is shown in correlation with the Ga depth scale. (c) ABF-STEM image of β-$Ga_2O_3$ sample irradiated with $1\times10^{16}$ $Ni/cm^2$. Panels (d) and (e) illustrate SAED patterns from the unimplanted and implanted regions respectively; (f-h) FFTs from high-resolution images taken from different regions of the sample as indicated in the panel (c).

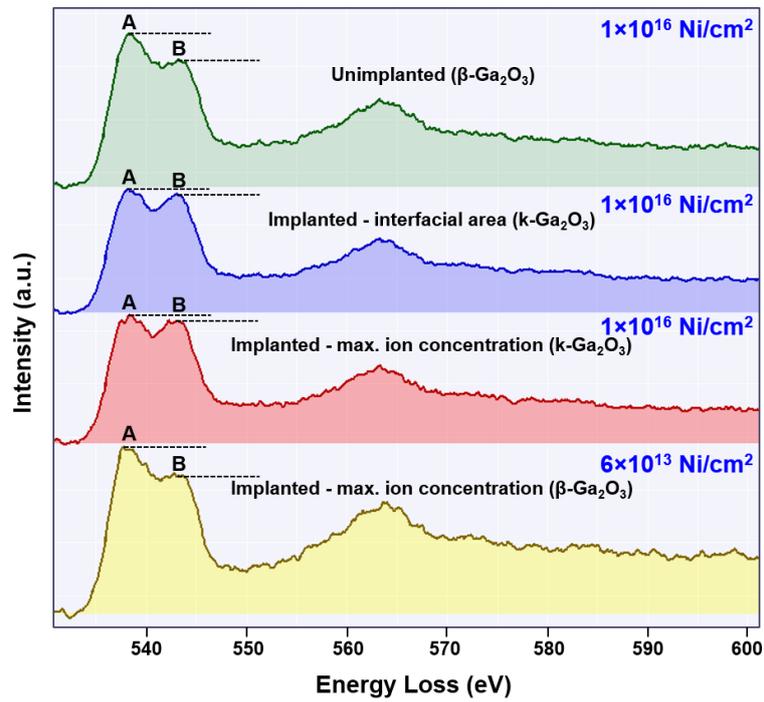

**Figure 2.** EELS spectra of the oxygen-K edge, acquired from different areas of the high-dose and low-dose implanted β-$Ga_2O_3$, illustrating characteristic intensity changes correlated with β-to-κ transition.

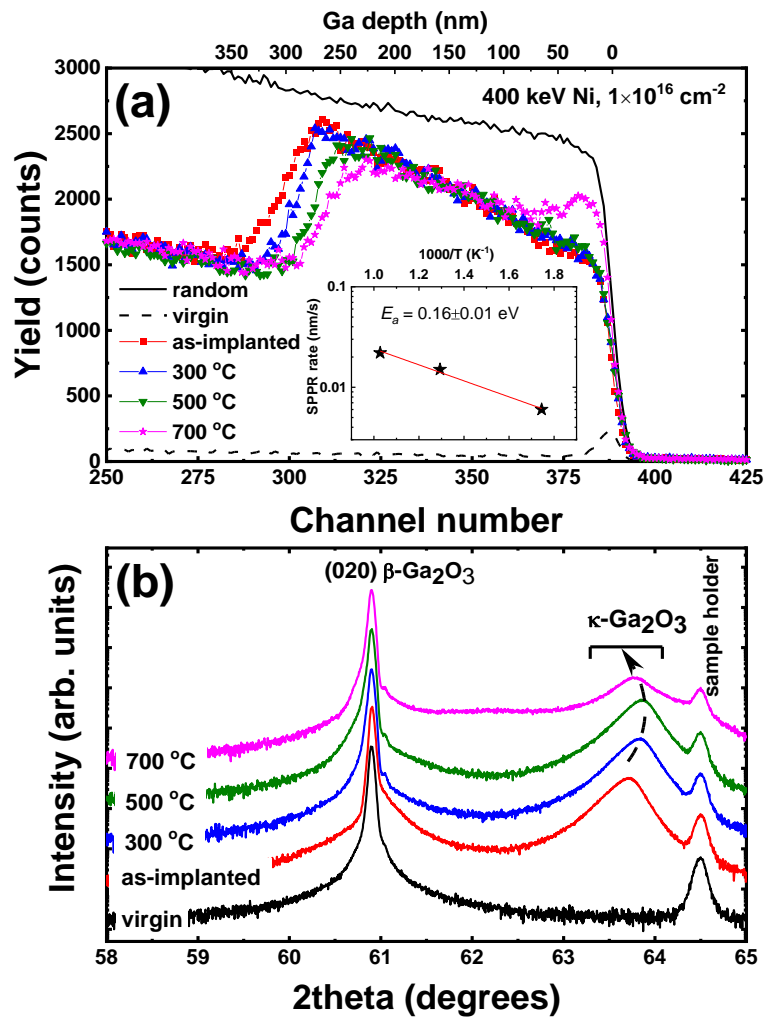

**Figure 3.** Thermal stability of the ion-beam-induced κ-Ga$_2$O$_3$ film. (a) RBS/C spectra and (b) corresponding XRD 2theta scans of the sample shown in Fig. 1(c) before and after isochronal anneals as indicated in the legend. The inset in the panel (a) shows Arrhenius plot of the solid-phase polymorphic regrowth rate as deduced from the RBS/C data.

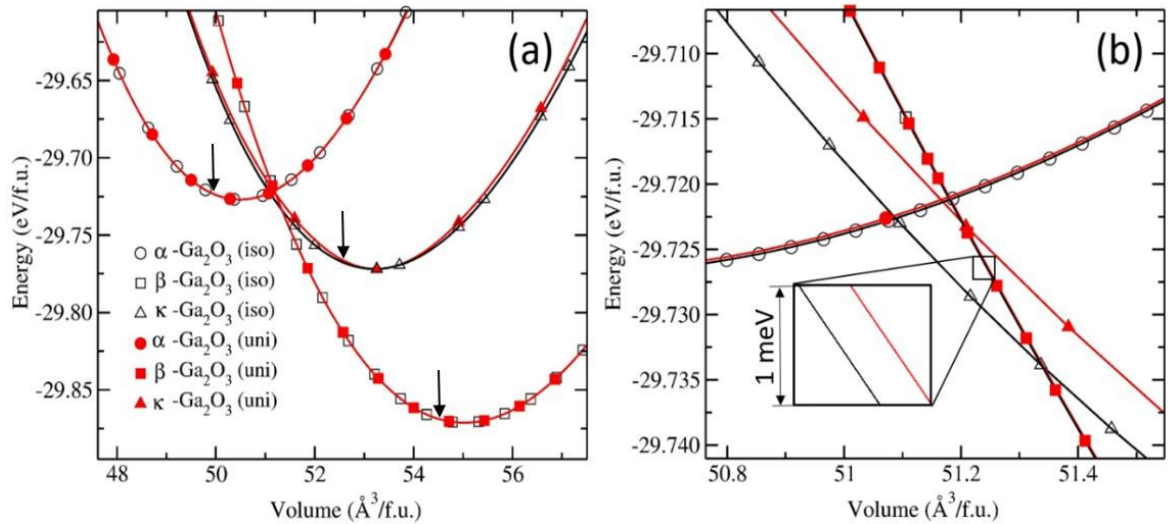

**Figure 4.** DFT calculated lattice energies for α-, β-, and κ-$Ga_2O_3$ as a function of the volume per formula unit for (a) bigger interval showing energy minima for the corresponding phases and (b) smaller interval emphasizing the transition points. The data were collected under isotropic (open symbols) and uniaxial stress along [010] direction (filled symbols). Experimental volumes of different phases are indicated by the arrows.

Supplementary Materials for

Disorder-induced ordering in gallium oxide polymorphs

Alexander Azarov[1,a], Calliope Bazioti[1], Vishnukanthan Venkatachalapathy[1,2],

Ponniah Vajeeston[3], Edouard Monakhov[1], and Andrej Kuznetsov[1, b]

[1] *University of Oslo, Department of Physics, Centre for Materials Science and Nanotechnology, PO Box 1048 Blindern, N-0316 Oslo, Norway*

[2] *Department of Materials Science, National Research Nuclear University, "MEPhI", 31 Kashirskoe Hwy, 115409 Moscow, Russian Federation*

[3] *University of Oslo, Department of Chemistry, Centre for Materials Science and Nanotechnology, PO Box 1033 Blindern, N-0315 Oslo, Norway*

Table of Contents

[a] alexander.azarov@smn.uio.no

[b] andrej.kuznetsov@fys.uio.no

## Section I. Identification of the κ-phase XRD peaks and anisotropy effects

Figure S1 shows 2theta scans of the (010) oriented β-Ga$_2$O$_3$ crystals implanted with Ni and Au ions in comparison with the virgin sample. Accounting for the SAED pattern in Fig. 1(e), the peak at 63.76° in the sample implanted with 1×10$^{16}$ Ni/cm$^2$ can be unambiguously identified; it is associated with the diffraction from the {010} family of planes and specifically (060) planes in κ-phase, see the corresponding dashed line label in Fig. S1. Notably, other signatures of the {010} planes were not observed, attributed to the structure factor exclusions.

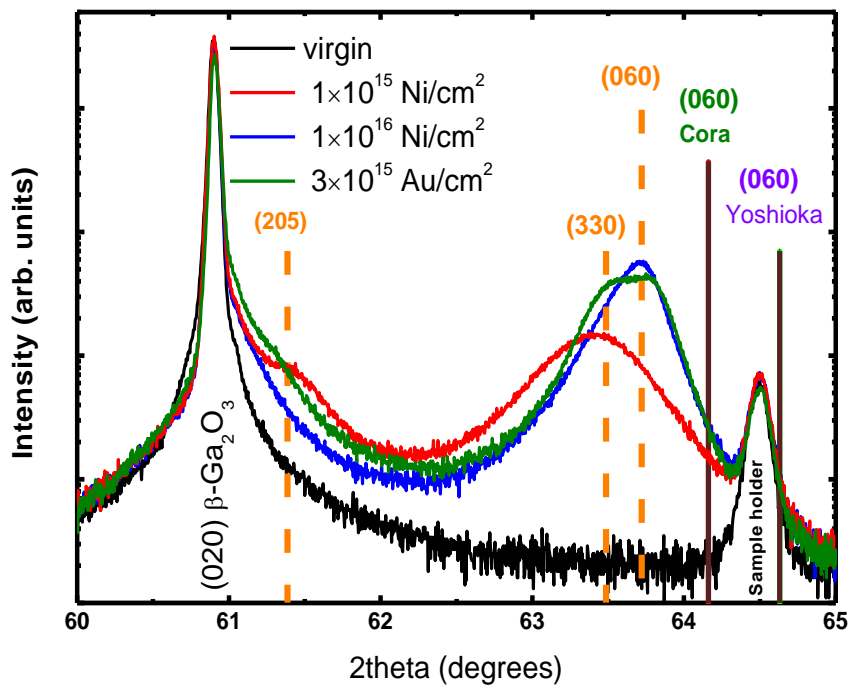

**Fig. S1** XRD 2theta scans of the (010) oriented β-Ga$_2$O$_3$ crystals implanted with 1×10$^{15}$ Ni/cm$^2$, 1×10$^{16}$ Ni/cm$^2$ and 3×10$^{15}$ Au/cm$^2$. The scan of the virgin β-Ga$_2$O$_3$ crystal is also included for comparison. Continuous and dashed lines label diffraction positions based on the literature [SR1,18] and the present analysis, respectively.

Notably, 63.76° for the κ-phase (060) diffraction is ~ 0.5° and 1° lower comparing with the data reported by Cora et al [SR1] and by Yoshioka et al [18], respectively, as also labelled by the corresponding continuous lines in Fig. S1. Converting the angular position to the interplanar distance, we obtain 0.1459 nm in comparison with 0.1450 nm by Cora et al [SR1]. Considering that 0.1450 nm characterizes the relaxed lattice, our data indicate +0.62% strained lattice, assuming isotropic strain. In the following, we will account for this order of strain as a correction factor while identifying other peaks.

The sample implanted with $3\times10^{15}$ Au/cm$^2$ was designed to mimic the ballistic defect production rates for that in $1\times10^{16}$ Ni/cm$^2$ and as such was expected to result into similar phase transformation. However, for this sample, a double peak centered at 63.52° and 63.76° is observed, see Fig. S1. The 63.76° peak can be readily assigned to the κ-phase (060) diffraction as discussed above, while the 63.52° peak can be associated with κ-phase (330) planes shifted from Cora et al data [SR1] by the strain factor of +0.61 %. Further, in the sample implanted with $1\times10^{15}$ Ni/cm$^2$, the (060) diffraction is missing, however the (330) related diffraction is still observed, even though it is centered at 63.43° - see Fig. S1 – consistently with the strain factor correction of +0.72 %. Altogether it suggests a very interesting dynamics of the β-to-κ phase transformation upon the accumulation of the radiation induced lattice disorder; exhibiting a balance in terms of the preferential orientation(s) as a function of the disorder accumulation. Indeed, the difference between $3\times10^{15}$ Au/cm$^2$ and $1\times10^{16}$ Ni/cm$^2$ implants may be attributed to the non-linear cascade density effects [SR3] for ions with significantly different masses (see also Supplementary Materials III). Lastly, in respect with Fig. S1, there is an additional peak, centered at 61.33° and most clearly seen in the sample implanted with $1\times10^{15}$ Ni/cm$^2$. This peak is attributed to the κ-phase (205) planes with the strain factor correction of +0.72 % in respect with Cora et al data [SR1]. Notably, the analysis above excludes contributions from planes belonging to other Ga$_2$O$_3$ polymorphs; primarily because we know from STEM that single κ-phase forms and additionally because all other polymorphs are higher in energy. In fact, considering the energetic aspects (see Supplementary Materials - IV), the only additional realistic option is to add α-phase into considerations here, but as we showed it above, the data in Fig. S1 are explainable in terms κ-phase only.

The implants done into (-201) oriented β-Ga$_2$O$_3$ crystals confirm the general character of the phenomena. Indeed, Fig. S2 shows XRD-2theta scans of the (-201) oriented β-Ga$_2$O$_3$ crystals implanted with Ni$^+$ ions (a) (-402) diffraction centered at 38.38° and (b) (-603) diffraction centered at 59.08°. For the low dose implant, the shoulder peak forms at the low-angle side of both (-402) and (-603) reflections. Notably, these shoulder peaks comprise the well-defined fringes, with increasing number of fringes from (-402) to (-603) reflections. This observation resembles that in Fig. 1(a), however the planes in β-Ga$_2$O$_3$ in Fig. S2 are subjected to the tensile strain. Meanwhile, similarly to that in Fig. 1(a), for the highest dose (1×10$^{16}$ Ni/cm$^2$), the strain is released and the new polymorph formation is observed with a double peak in Fig. S2(a) centered at 37.75° and 36.67°. In the first approximation, neglecting the anisotropy effects, the same implantation dose might produce the same radiation-induced disorder, in its turn resulting into the same β-to-κ phase transformation in both (010) and (-201) oriented β-Ga$_2$O$_3$ crystals. However, evidently, the anisotropy plays a role, resulting in a double XRD peak in Fig. S2(a) explainable either in terms of the double orientation or double phase. Also very interestingly, 2×10$^{14}$ Ni/cm$^2$ implants were still associated with the strain accumulation in the (010) oriented β-Ga$_2$O$_3$ – see Fig. 1(a); however, distinct diffraction peak at 36.67° forms for the same implantation conditions in (-201) oriented β-Ga$_2$O$_3$ – see Fig. S2(a), confirming the importance of the anisotropy effects. Nevertheless, qualitatively, the trend for β-to-κ transformation holds. Applying 1.75% compressive strain correction for the planes positions from literature [SR1], we may label (201), (131) and (210) κ-phase planes at 36.88°, 37.00°, and 37.07°, respectively, see Fig. S2. Notably, there could be alternative identification scenarios for peaks in Fig. S2(a) proposed in terms of other polymorphs too; e.g. ε-$Ga_2O_3$, γ-$Ga_2O_3$ and δ-$Ga_2O_3$ exhibit peaks centered at 37.14°, 37.27° and 38.28° associated with (101), (222) and (332) planes, respectively. However, as mentioned above, these options were excluded from the consideration, because of the STEM observation and the energetic reasons.

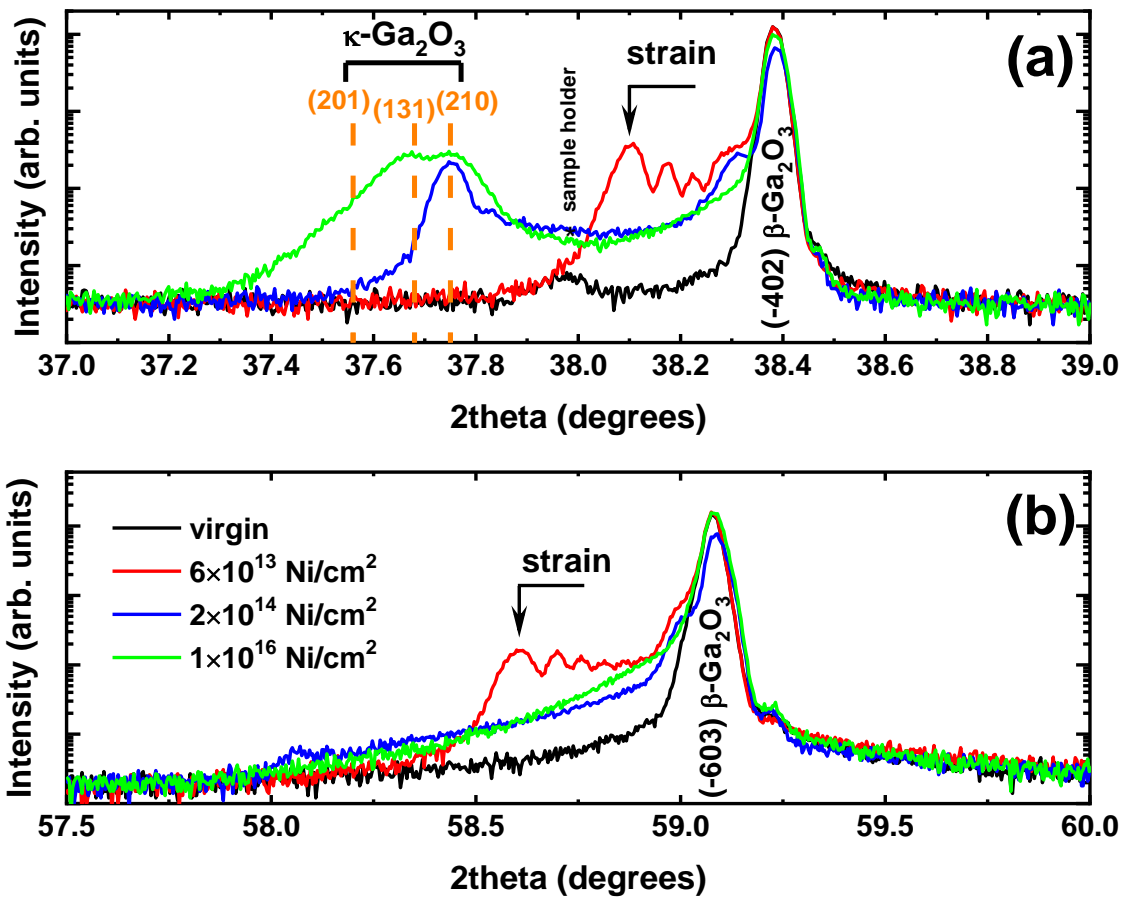

**Fig. S2** XRD 2theta scans across (-402) and (-603) reflections of the (-201) oriented β-$Ga_2O_3$ crystals implanted with $Ni^+$ ions as indicated in the legend. Dashed lines label the κ-$Ga_2O_3$ planes responsible for the diffraction based on present analysis.

Notably, no new phase-related peaks appear in Fig. S2(b) around (-603) reflection of β-$Ga_2O_3$. The broader background observed for (-603) reflection of β-$Ga_2O_3$ may be attributed to the extended defects or lattice distortions [SR4].

## Section II. Analysis of the adjacent κ-phase grains exhibiting different strain contrast.

Even though the film in Fig. 1(c) was clearly interpreted as single κ-phase film, there are still remaining questions, in particular related to potential mis-orientation between the grains as well as regarding potential chemical variations. For that reason, we investigated two adjacent grains in Fig. S3. As seen from Fig. S3, the grains are nicely co-oriented and there are no chemical inhomogeneities observed.

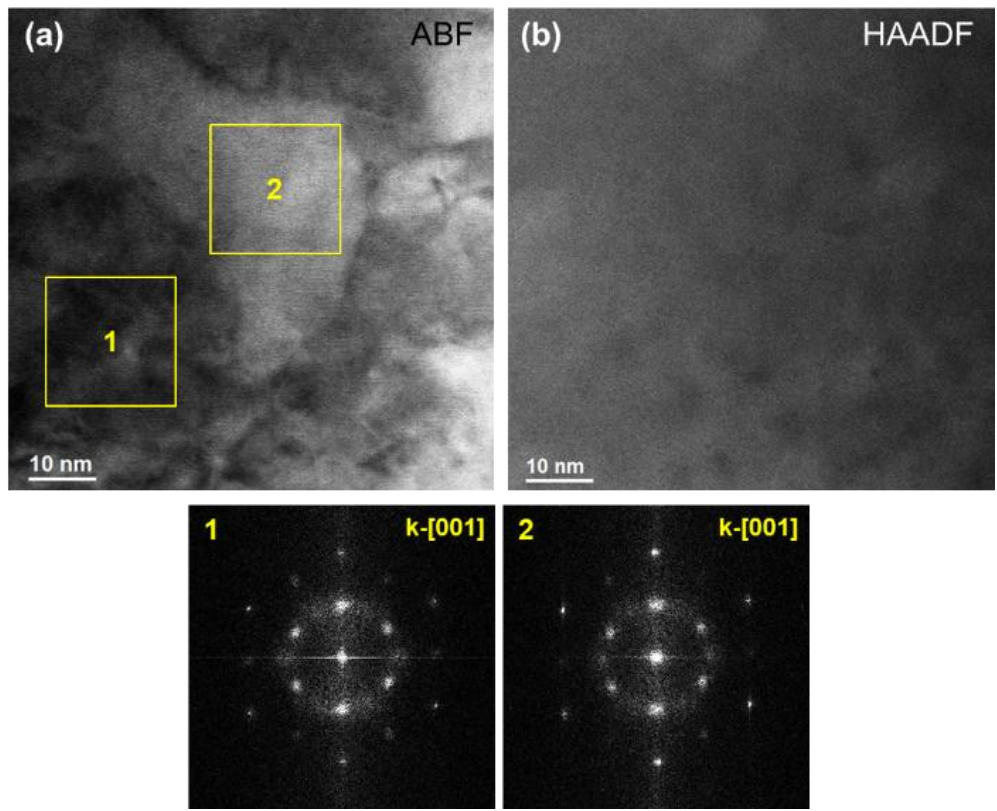

**Fig. S3** (a) ABF-STEM and (b) HAADF-STEM high-resolution images acquired simultaneously from the maximum ion-concentration region of (010) β-Ga$_2$O$_3$ irradiated with $1\times10^{16}$ Ni/cm$^2$. FFTs extracted from two adjacent grains reveal that both grains are stabilized in κ-phase and have the same orientation. The stable contrast in HAADF (pure Z-contrast image) indicates no chemical variations and the contrast in ABF is attributed only to the strain.

### Section III. Microstructure of the low-dose implanted β-Ga$_2$O$_3$ sample

Fig. S4 shows the STEM analysis of the sample implanted with 6×10$^{13}$ Ni/cm$^2$ in panels (a-d). The ADF-STEM image reveals that the strain contrast in the implanted region gradually reduces when moving away from the $R_{pd}$, with no signs of abrupt changes, see Fig. S4(a). This region is likely to contain point defects and defect clusters, but no amorphous areas were detected in the entire implanted region. This is confirmed by a combination of the SAED and FFT in Figs. S4(b-d). Thus, β-phase is identified through all regions in Fig. S4(a) with no signs of phase transitions; meanwhile the accumulation of strain is confirmed in this sample.

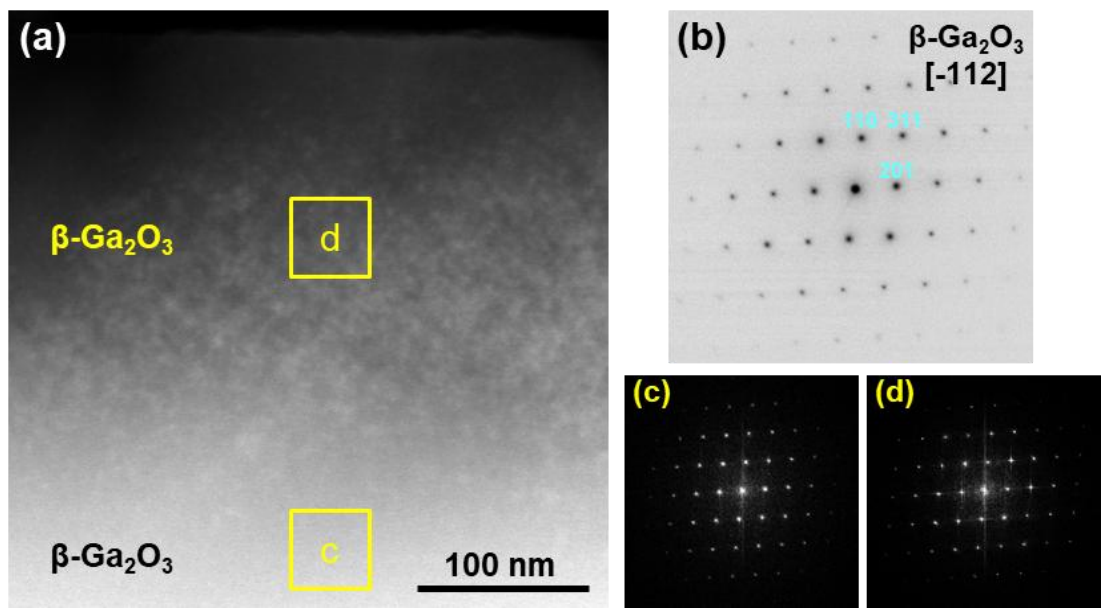

**Fig. S4** Microstructure of the low-dose implanted β-Ga$_2$O$_3$ sample (a) ADF-STEM image of β-Ga$_2$O$_3$ sample implanted with 6×10$^{13}$ Ni/cm$^2$, (b) corresponding SAED pattern at the depth of $R_{pd}$ and (c)-(d) FFTs from high-resolution images taken from the unimplanted and implanted regions, respectively.

## Section IV. Implants with different ions normalized to produce similar number of atomic displacements

In the main part of the paper we selected to show the data for systematic $^{58}$Ni$^+$ ion implants, however $^{69}$Ga$^+$, and $^{197}$Au$^+$ implants data were systematically exploited too, and used for comparison to demonstrate the general character of the phenomena (as already shown in Fig. S1 in Supplementary Materials – I). For that reason, the ballistic defect production rates (without accounting for non-linear cascade density effects [SR3]) for $^{69}$Ga$^+$, and $^{197}$Au$^+$ implants were normalized to that of $^{58}$Ni$^+$ ion implanted with 400 keV in a wide dose range of 6×10$^{13}$ - 1×10$^{16}$ cm$^{-2}$. Specifically, we used SRIM code [SR5] simulations to choose roughly similar profiles of the primary defects or displacements per atom (DPA), considering similar approach as that used earlier by Titov et.al. [19]. In our calculations, displacement energies of 25 and 28 eV were used for Ga and O atoms, respectively. Fig. S5 provides a representative example of the corresponding RBS/C and XRD data, while the inset in Fig. S5(a) showing the DPA depth profiles of DPA for Ni, Ga and Au implants. As seen from Fig. S5, both the RBS/C profiles and XRD 2theta scans exhibit very similar trends for the all ion species used. Thus, the observed polymorph transitions are attributed to the disorder-induced effects with negligible impact of the chemical nature of the ions. Notably, Au ions produced the thickest modified layer, as seen from Fig. S5(a), despite that the corresponding DPA profile is most shallow, see the inset in Fig. S5(a). This effect is readily explainable in terms of variations in the cascade density effects [SR3, 19] not taken in account by SRIM.

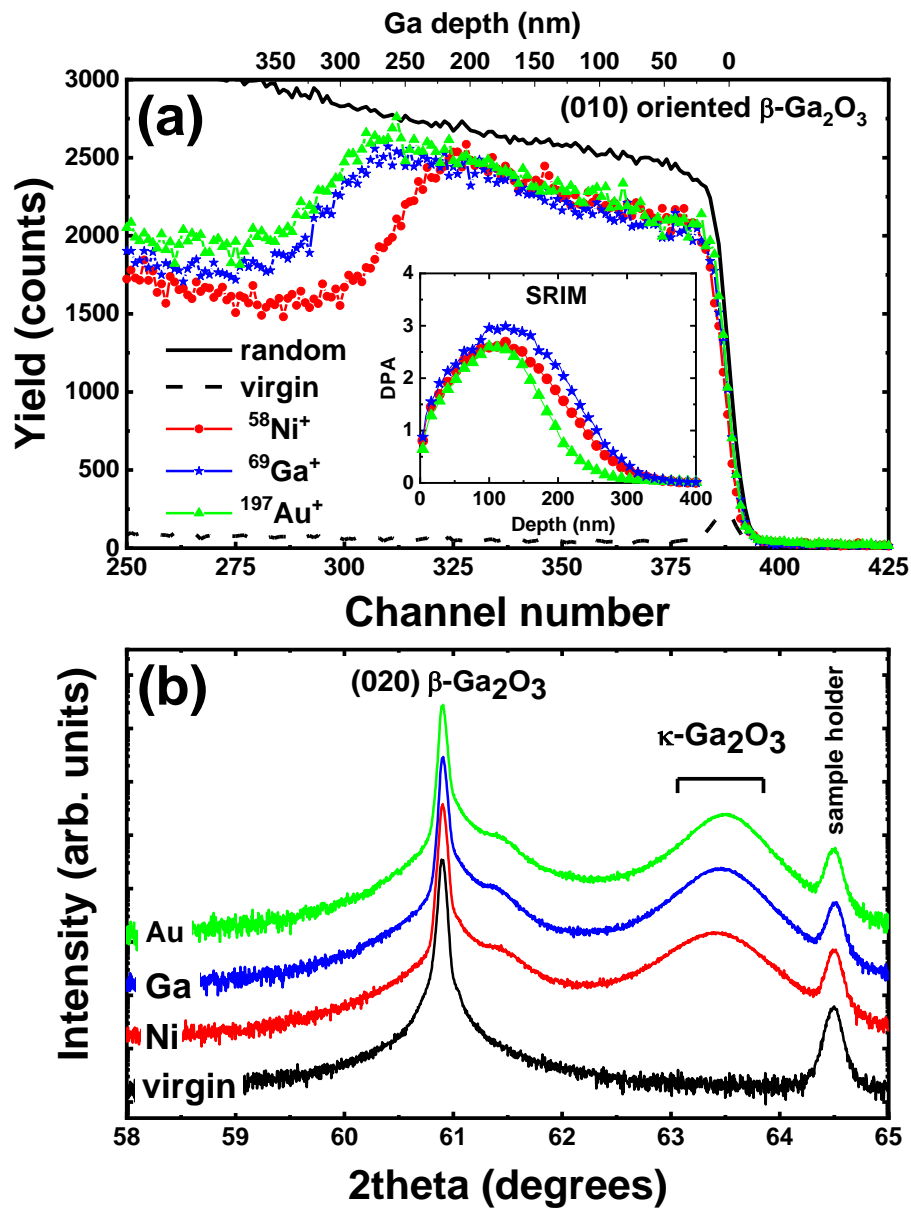

**Fig. S5** (a) RBS/C spectra and (b) XRD 2theta scans of the (010) oriented β-Ga$_2$O$_3$ crystals implanted with 400 keV $^{58}$Ni$^+$ (1×10$^{15}$ cm$^{-2}$), 500 keV $^{69}$Ga$^+$ (1×10$^{15}$ cm$^{-2}$) and 1.2 MeV $^{197}$Au$^+$ ions (3×10$^{14}$ cm$^{-2}$). The inset in the panel (a) shows the depth profiles of displacements per atom (DPA) as generated by Ni, Ga and Au ions ion impacts, as simulated by the SRIM code [SR5].

## Section V. Geometric phase analysis of the interfacial area

In order to unravel the strain-state of the κ/β-Ga$_2$O$_3$ interface, nanoscale strain analysis was conducted by applying Geometric Phase Analysis (GPA) of the high-resolution TEM images, see Fig.S6. To start with, Fig. S6(a) shows the high-resolution TEM image of the interfacial region in the high-dose implanted (1×10$^{16}$ Ni/cm$^2$) sample. The sample was oriented along the [102] zone axis of the β-phase and the orientation relationship between two phases at the interface is described by:

(20-1)$^β$//(002)$^κ$ (in-plane matching, planes vertical to the interface)

(020)$^β$//(060)$^κ$ (out-of plane, planes parallel to the interface)

Further, for all panels in Fig.S6, the yellow dashed lines select the analysis region, with yellow arrows pointing at the κ/β interface. Figs. S6(b) and (c) illustrate the extracted $ε_{xx}$ (in-plane) and $ε_{yy}$ (out-of-plane) strain maps, respectively. Figs. S6(d) and (e) plot the corresponding lattice strain profiles, depicting the relative deviations of the inter-planar spacing (*d*) with respect to that of the β-Ga$_2$O$_3$ substrate (reference area). Notably, the black arrows in panels (b-c) indicate the direction of strain profiles in panels (d-e). Given that the nominal *d* for each phase are $d^{(20-1)β}$ = 0.4679 nm, $d^{(002)κ}$ = 0.4641 nm, $d^{(020)β}$ = 0.1518 nm and $d^{(060)κ}$ = 0.1450 nm [SR1], the nominal in-plane ($ε_{xx}$) and out-of-plane ($ε_{yy}$) lattice differences for the relaxed phases are:

$$ε_{xx} = \frac{d(002)κ - d(20-1)β}{d(20-1)β} * 100 = -0.81\%$$

$$ε_{yy} = \frac{d(060)κ - d(020)β}{d(020)β} * 100 = -4.49\%$$

Thus, the GPA analysis shows the formation of two strained interfacial zones of equal thickness ~4 nm each. The first zone forms in the upper part of the β-Ga$_2$O$_3$ area before the phase transition took place, characterized with the compressive out-of-plane strain of -1%. In contrast, the interfacial section of the κ-phase is found to be stretched (+2.5 % in-plane, +1.25 % out-of-plane) forming the second zone.

Phys. Rev. Lett. **128**, 015704 (2022)
Doi: https://doi.org/10.1103/PhysRevLett.128.015704

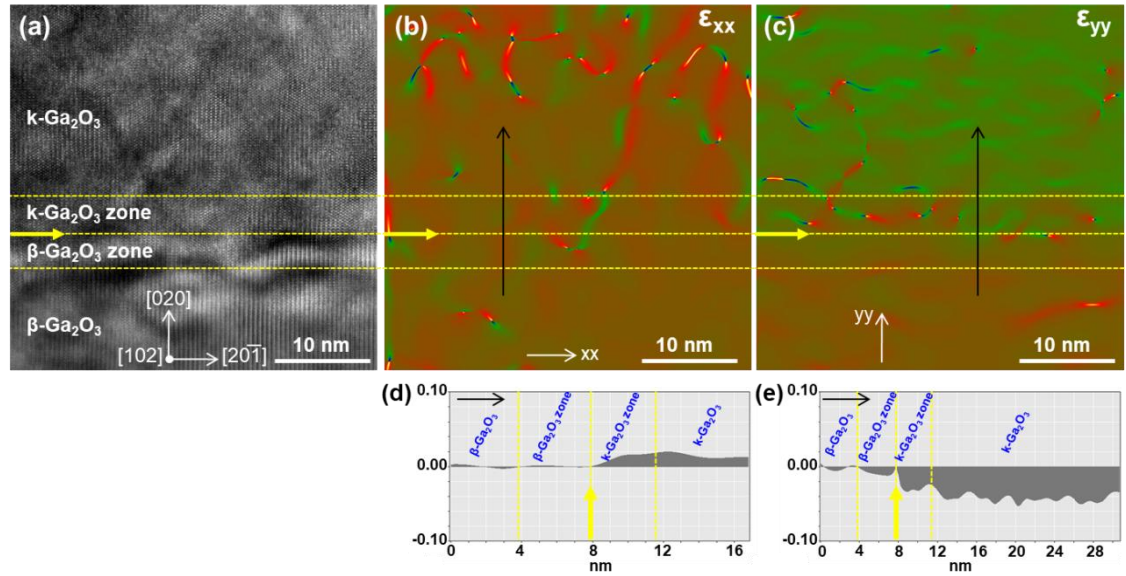

**Fig. S6** (a) High-resolution TEM image showing the interfacial area of the high-dose implanted ($1\times10^{16}$ Ni/cm$^2$) sample, (b)-(e) the $\varepsilon_{xx}$ and $\varepsilon_{yy}$ strain maps and the corresponding strain profiles, respectively, as indicated in the figure. For all panels, the dashed yellow lines select the analysis region, with yellow arrows pointing at the κ/β interface. The black arrows in panels (b-c) indicate the direction of strain profiles.

These strained zones are the remnants of the β to κ phase-transition and are likely the most vulnerable areas to recrystallize back to β-phase when thermal annealing occurs. More far away from this interfacial strained κ-Ga$_2$O$_3$ zone, the κ-Ga$_2$O$_3$ phase starts to relax, with the values being reduced to +2.1 % (in-plane) and +0.62 % (out-of-plane). This is in full accordance with the XRD results, showing a shift of the (060)$^\kappa$ planes indicating +0.62 % out-of-plane strain (Fig. S1). For convenience, we sum up the results of the GPA shown in to the table below:

| region | $\varepsilon_{xx}$ | $\varepsilon_{yy}$ | calculated d (nm) | strain-state |
|---|---|---|---|---|
| β-Ga$_2$O$_3$ (ref.) | 0 | 0 | $d^{(20\text{-}1)\beta}$ = 0.4679 nm $d^{(020)\beta}$ = 0.1518 nm | relaxed |
| β-Ga$_2$O$_3$ zone 4 nm | 0 | -0.01 | $d^{(20\text{-}1)\beta}$ = 0.4679 nm $d^{(020)\beta}$ = 0.1503 nm | yy -1% |
| κ-Ga$_2$O$_3$ zone 4 nm | +0.017 | -0.03 | $d^{(002)\kappa}$ = 0.4759 nm $d^{(060)\kappa}$ = 0.1468 nm | xx +2.5% yy +1.25% |
| κ-Ga$_2$O$_3$ | +0.012 | -0.039 | $d^{(002)\kappa}$ = 0.4738 nm $d^{(060)\kappa}$ = 0.1459 nm | xx +2.1% yy +0.62% |

## Section VI. Ga$_2$O$_3$ polymorphs equilibrium and pressure induced transformations

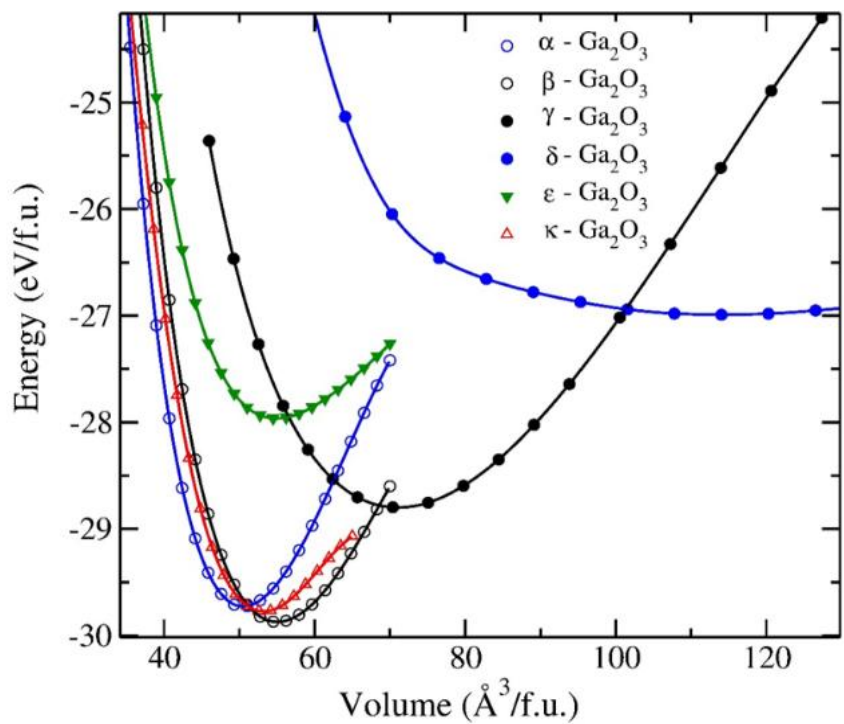

**Fig. S7** Lattice total energy vs volume per formula unit for six different Ga$_2$O$_3$ polymorphs obtained with DFT calculations.

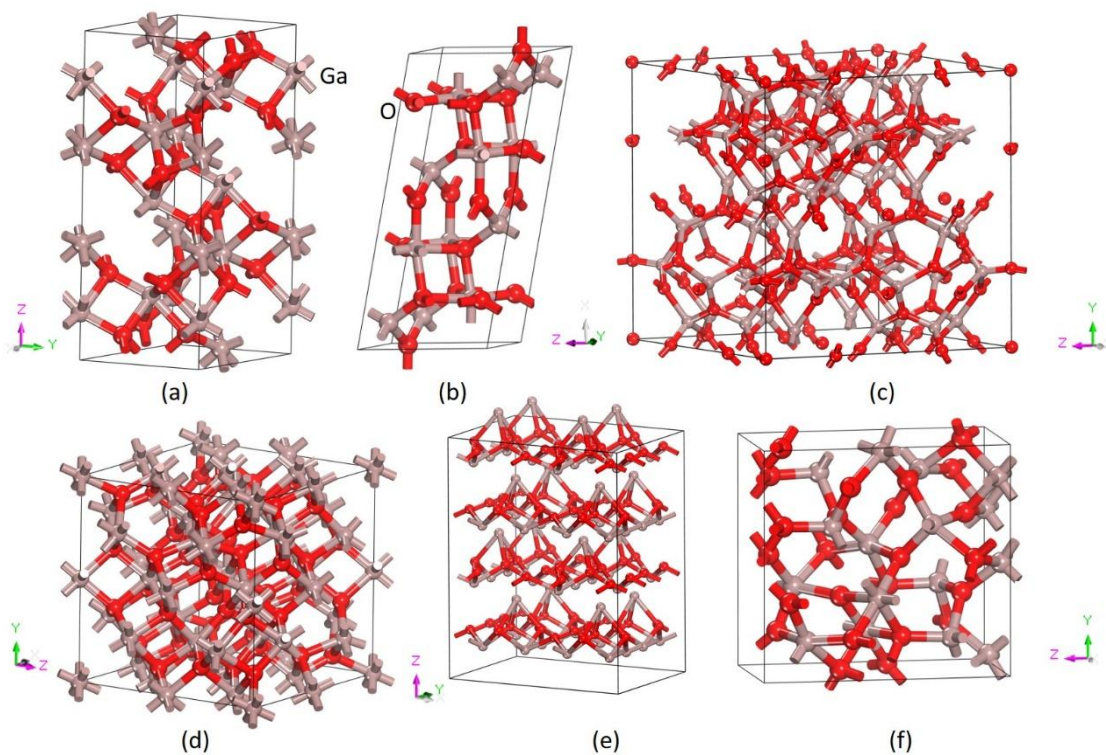

**Fig. S8** Theoretically optimized crystal structures for (a) α-, (b) β-, (c) γ-, (d) δ-, (e) ε- and (f) κ-Ga$_2$O$_3$ structures at equilibrium volumes. Ga and O atoms are color-discriminated. Notably, x, y, and x correspond to [100] [010], and [001] directions, respectively.

In the present work, we consider six experimentally determined polymorphs (α, β, κ, δ, γ and ε) of $Ga_2O_3$ and Fig. S7 summarizes DFT calculated the total energy vs volume per formula unit for these $Ga_2O_3$ polymorphs. The visual representation of the corresponding $Ga_2O_3$ polymorphs is given in Fig. S8. The frozen phonon calculation was performed on these polymorphs using the Phonopy program to obtain the phonon dispersion curve and the density of states (DOS) [20]. An atomic displacement of 0.0075 Å was used, with a symmetry consideration, to obtain the force constants for the phonon calculations. The displacements in opposite directions along all axes were incorporated in the calculations to improve the overall precision. The force calculations were made using the VASP code with the supercell approach and the resulting data were imported into the Phonopy program. The dynamical matrices were calculated from the force constants, and phonon DOS curves were computed using the Monkhorst-Pack scheme [21]. As an example, Fig. S9 shows the calculated total phonon DOS for α-, β-, and κ-polymorphs confirming their dynamic stability.

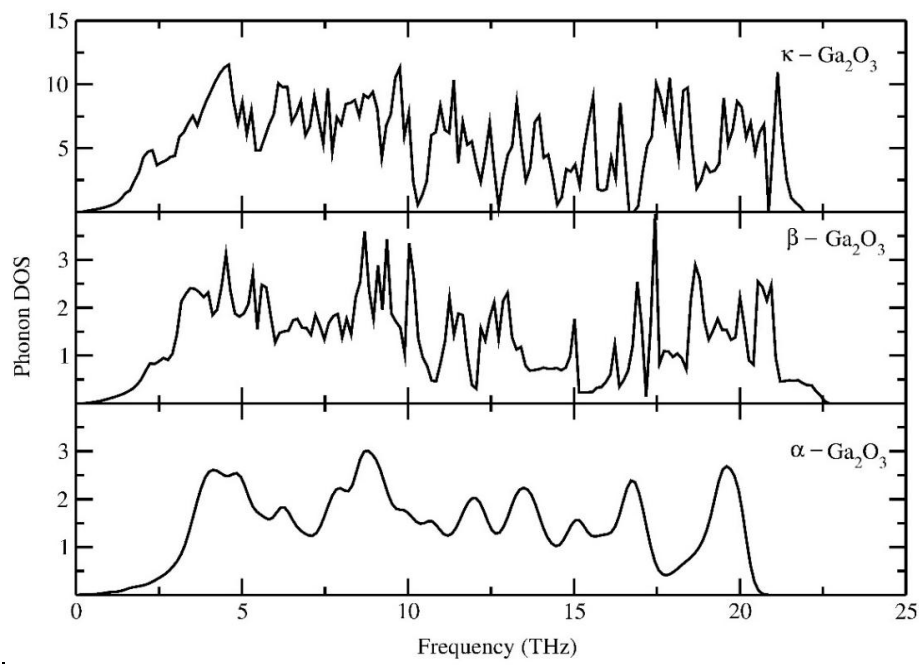

**Fig. S9** Calculated phonon density of states for α-, β-, and κ-polymorphs of $Ga_2O_3$ at 0 GPa pressure and 0 K temperature.

Further, the effect of the applied external pressure was investigated for α-, β-, and κ-polymorphs and Fig. S10 illustrate the situation setting the Gibbs energy of the β-phase to zero. Fig. S10 reveals that the β- to-κ phase transition occurs at the range of 9.8 - 13.4 GPa accounting for the isotropic pressure. This is in agreement with the observation of the β-to-α phase transition for pressures exceeding 13.6 GPa [22]. Importantly, the actual energy difference between α-, β-, and κ-polymorphs is fairly small – as illustrated by Fig. 4 in the main part of the paper – providing an opportunity for the structure to remain in the metastable form even after the pressure release. Notably, the application of the isotropic pressure may be regarded as naïve. Indeed, β- $Ga_2O_3$ has the compressibility for the *a*, *b*, and *c* lattice-parameters of 6.0, 3.7, and 4.2 %, respectively, and β-angle increased by 1.9%; α- $Ga_2O_3$ had the compressibility of the *a* and *c* lattice-parameters of 3.3 and 7.1 %, respectively [23]. In practice, different "threshold" pressures for $Ga_2O_3$ polymorph transitions were reported experimentally [SR11]. In our data, we also noticed that different levels of disorder, and subsequently of strain, are required for igniting the β- to-κ phase transition in differently oriented β- $Ga_2O_3$ crystals (see e.g. Figs. S1 - S2). Thus, we have also tested the application of the uniaxial pressure as shown in Fig. 4 in the main text of the paper.

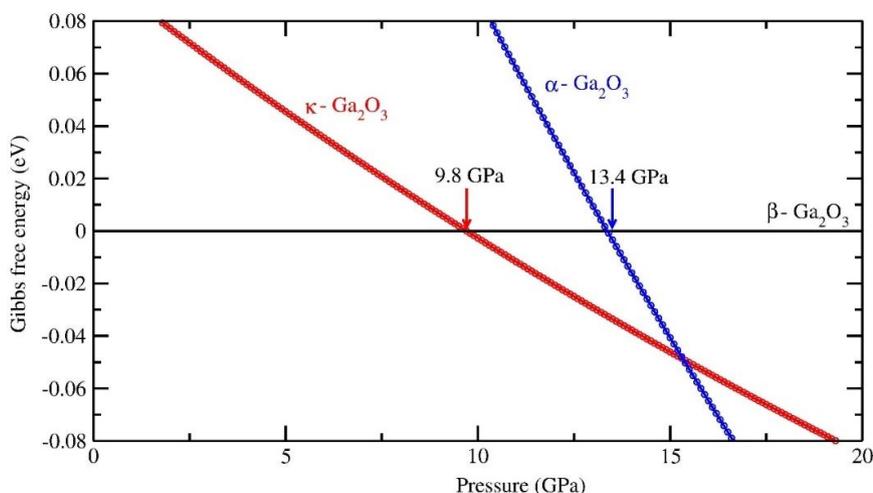

**Fig. S10** Stability of κ and α-$Ga_2O_3$ polymorphs relative to β-$Ga_2O_3$ as a function of the isotropic pressure. The transition pressures are marked by arrows at the corresponding transition points.

In addition to the anisotropic effects, temperature can have a decisive role in controlling the direction of the metastable strain-induced polymorph transitions. The free energies including entropic terms (*S*) were calculated accounting for the phonon simulation data, given as a function of temperature (*T*) by $G(T,p) = E^{tot} + E^{ZPE} - TS^{vib} + pV$. Where, the total energy ($E^{tot}$) is essentially plotted in Fig. S7; $E^{ZPE}$ is a zero point energy obtained from total phonon density; and *pV* is the pressure-volume factor [20].

Fig. S11 shows the free energy for β-, α- and κ-polymorphs as a function of temperature at 0 and 10 GPa pressures, showing that the temperature variations will obviously increase the complexity; however, potentially leading to an additional flexibility to obtain single-phase κ-$Ga_2O_3$, instead of the κ/α mixture. Indeed, as seen from Fig. S11, the energy difference between α- and κ-phases increases with increasing temperature for the sample subjected to pressure. This effect is evident already at room temperature and extends through the whole κ-phase stability region as we observed it in Fig. 3 in the main part of the paper.

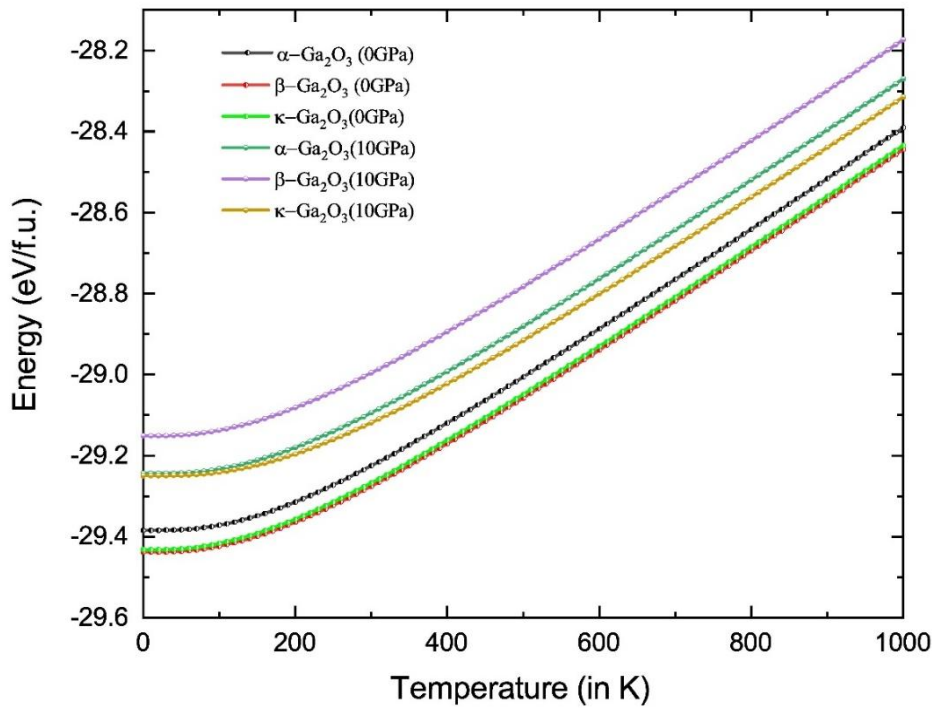

**Fig. S11** Calculated Gibbs free energy for $Ga_2O_3$ polymorphs as a function of temperature 0 GPa and at 10 GPa.

Phys. Rev. Lett. **128**, 015704 (2022)
Doi: https://doi.org/10.1103/PhysRevLett.128.015704

## Section VII. Effect of the implantation temperature

As predicted by Fig. S11, the energy difference between α- and κ-phases increases with increasing temperature for the sample subjected to pressure. However, temperature might also affect the balance between out-diffusion and annihilation of the radiation-induced defects, making it to an important factor for the localization of the β/κ interface in respect with the $R_{pd}$ region, as illustrated in Fig. S12. As seen from Fig. S12(a), the film thickness shrinks as a function of the implantation temperature, while Fig. S12(b) confirms the maintenance of the κ-phase XRD-peaks, evolving to a double-peak for the 300 °C implant.

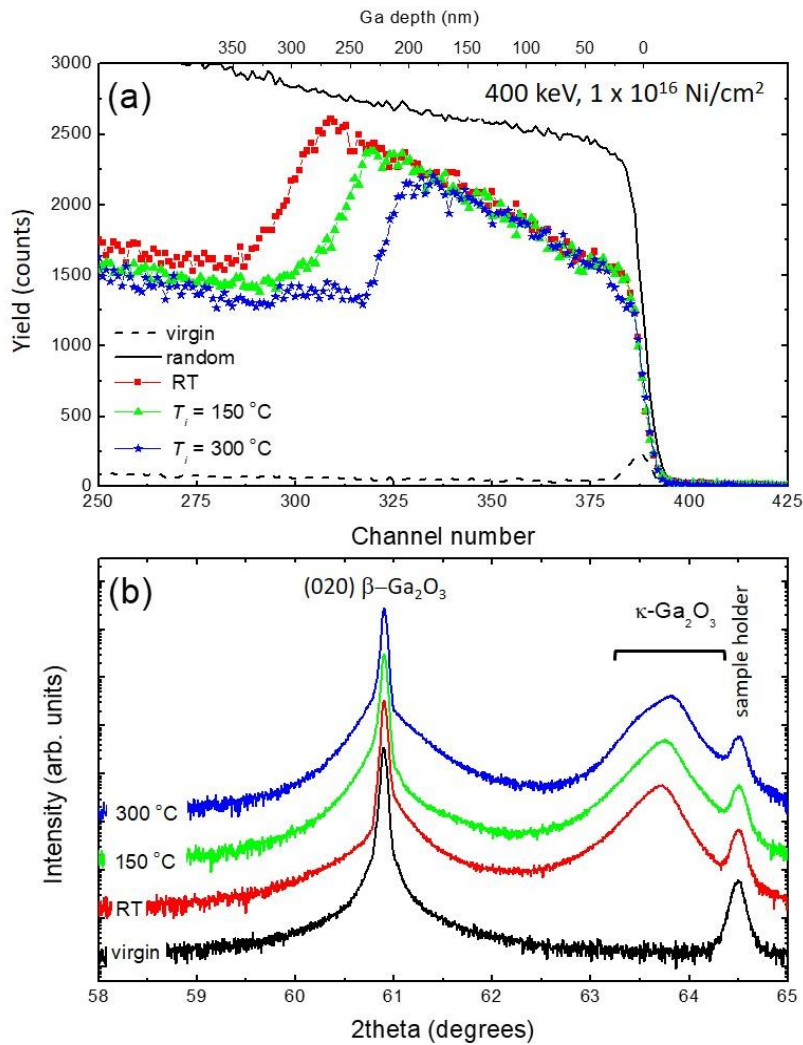

**Fig. S12** (a) RBS/C spectra and (b) XRD 2theta scans of the (010) oriented β-$Ga_2O_3$ crystals implanted with 400 keV, $1\times10^{16}$ Ni/$cm^2$ at different temperatures as indicated in the legend. The data for the room temperature implant are the same as in Figs. 1 and 3.